\documentstyle[preprint,aps,prb,psfig]{revtex} \makeatletter
\def\eqalign#1{\null\vcenter{\def\\{\cr}\openup\jot\m@th
  \ialign{\strut$\displaystyle{##}$\hfil&$\displaystyle{{}##}$\hfil
      \crcr#1\crcr}}\,} \makeatother

\begin{document} 

  \title{Non-equilibrium steady state of  the superfluid mixtures of helium isotopes at classical temperatures}

\author{A.I. Chervanyov}

 \address{ Max-Planck-Institut f\"ur Physik komplexer Systeme\\ N\"othnitzer
Str. 38, D-01187, Dresden, Germany\\ E-mail: chervany@milou.mpipks-dresden.mpg.de\\ \medskip and   Kharkov State University, 4 Svoboda  Sq., Kharkov 310077, Ukraine\\ E-mail: alex@viola.kharkov.ua}
\maketitle

  \begin{abstract}
 The size effect in the gas of impuritons of the superfluid
mixtures of helium isotopes is investigated by taking into consideration
the contribution of thermal excitations. The solution is obtained for  the set of kinetic equations describing a non-equilibrium state of the phonon-impuriton
system of a superfluid mixture situated in the volume filled with the  macroparticles. It allows to  find  the condition describing a  steady, thermodynamically  non-equilibrium state of $^3{He}-^4{He}$ mixture in confined geometry. The  Knudsen effect in the gas of impuritons of a superfluid mixture is investigated by
taking into account the  contribution of phonons. A  model for the  collision
operator has been proposed  to analyze the exact results 
in the context of the  concrete physical situations.  An  experiment for the  investigation of  the Knudsen effect in a superfluid mixture of helium isotopes is proposed.
\end{abstract}
  
\section{Introduction}

	There is a row of the  unique  features of a  size effect in the superfluid mixtures of helium isotopes. The first evident difference between the latter one and  it's classical analogue  is concerned with a non-ideality of the gases of quasiparticles of a superfluid mixture. It can be observed explicitly in the  degeneracy region  where the contribution of impuriton interaction to all the thermodynamic quantities  of the mixture reaches the significant values  
 ~[\ref{177}]. The approximation based on the classical Knudsen relation  
~[\ref{2}] can not be used  in the ultra low temperature region where the gas of impuritons shows it's quantum properties ~[\ref{377}].

	A very different situation takes place in a region of classical temperatures. Here, the contribution of quasiparticle interaction is very small  
~[\ref{177}] and  can not be considered as a reason for a deviation from the  well-known Knudsen relation.  On the other hand, the latter one, evidently, can  not be  used to describe the steady state of superfluid mixture even  at classical temperatures. Indeed it should be expected   that  a significant contribution of thermal excitations to the pressure  at the temperatures $T\succeq 0.6K$  ~[\ref{4}] results in the change of the  classical relation between the gradients of temperature and pressure describing the steady state of a  superfluid mixture.
	
	  The relation which describes the steady, thermodynamically non-equilibrium state of a superfluid mixture in confined geometry is  presented in this work . We use  the kinetic  method to investigate the  steady non-equilibrium state of a  superfluid mixture situated in the volume filled with a gas of macroparticles. Such an approach can be considered as a development of the  well-known "dust laden " gas model ~[\ref{577},\ref{6},\ref{7}]  describing the process of diffusion of the  "light" particles in a gas of "heavy" immobile macrons. It should be noted that the requirement of immobility  of the "heavy" particles in their collisions with the "light" ones should  be considered as an idealization. Indeed, it corresponds to the  zeroth order approximation  with respect to the ratio of the particle mass to the macron  mass ~[\ref{2}]. The model which is  considered  in this paper  is valid  for a broad  spectrum of practical applications. It can be  used to describe the process of particle diffusion in very different physical systems such as a volume filled with the crocus ~[\ref{Ebner}] and a gas of the  crystal formations associated with the  positive ions in a superfluid mixture  ~[\ref{8}].
 
	The paper is organized as follows: Section ${\bf II}$ presents the mathematical  solution of the  above described  problem  in the framework of the  exact kinetic approach. In Section ${\bf III}$  a convenient model for the inter-particle collision operator is proposed to investigate   the general solution of the kinetic equation and to obtain explicitly the mass velocities of the quasiparticle gases. The conditions of a steady state formation are obtained and analyzed in Section ${\bf IV}$  for  different relations between the frequencies of the  inter-quasiparticles and quasiparticles-macrons collisions. In this Section we also describe  an experiment which would make it possible  to check the obtained theoretical results. A brief summary is given in Section ${\bf V}$.

\section{Solution of the kinetic equations} 

	The approach based on the quasiparticle description ~[\ref{2}]  makes it possible to use the kinetic method for investigation of a slightly non-equilibrium state of a superfluid mixture of helium isotopes.  The specificity of the $^3He-^4He$  superfluid mixture externalizes in  the fact that one has to consider a non-equilibrium state of a three-component system of the  gases  of quasiparticles   ( phonons, rotons and impuritons ) with non-quadratic energy-momentum relations. If temperature is less than 0.9 K the contribution of rotons  to  all the thermodynamic quantities can be neglected and one can restrict  oneself to the 
analysis of the two -component system of the  impuritons and phonons. In order  to formulate exactly the kinetic problem it is necessary to  consider the Hilbert spaces $\Re _i$ of the momentum functions
$g\left( \overrightarrow{p_i}\right)$ 
 with the scalar product 

\begin{equation}
\label{sc1}
\left( h\left( \overrightarrow{p_i}\right) ,g\left( \overrightarrow{p_i}\right)\right) _i=-\int f_0^{\left( i\right) ^{\prime }}\left( \frac{\varepsilon _i}T
\right) h^{*}\left( \overrightarrow{p_i}\right) g\left(\overrightarrow{p_i}
\right) d\Gamma _i
\end{equation}
Here  $\overrightarrow{p_i}$  is the quasiparticle momentum in s-system ( the coordinate  system in which  a superfluid component is at rest), $d\Gamma _i$ is the corresponding volume element of the momentum phase space ( $i=3$  corresponds to impuritons, $i=4$ to phonons), $\mu _3$ is the  chemical potential of  the  impuritons, $f_0^{\left( 3\right) }\left( x\right) =\frac 1{\exp {\left( x-\frac{\mu _3}{T}\right)} +1}$,  $f_0^{\left( 4\right) }\left( x\right) =\frac 1{\exp(x)-1}$,  $\varepsilon _3=\frac{p_3^2}{2m^{*}}$, $\varepsilon_4=cp_4$, $c$ is the sound velocity, $p_i$ are the moduli of corresponding vectors,
the prime  denotes  differentiation  with respect to the  argument.

	We introduce  the Hilbert space $\Re$   of  vectors $\left| \phi\left( \overrightarrow{p_3}\right) ;\varphi \left( \overrightarrow{p_4}\right)\right\rangle ^T$
 with the components  $\phi\left( \overrightarrow{p_3}\right) $, $\varphi \left( \overrightarrow{p_4}\right)$  belonging to the  Hilbert spaces $\Re _3$ ,  $\Re _4$  respectively. The scalar product in the space  $\Re $  is defined by the equality:
\begin{equation}
\label{sc2}
\left\langle \phi _1\left( \overrightarrow{p_3}\right)\\; \varphi _1\left( \overrightarrow{p_4}\right) \left| 
\phi _2\left( \overrightarrow{p_3}\right)  \\; \varphi _2\left(
\overrightarrow{p_4}\right)\right. \right\rangle =
\left( \phi _1\left( \overrightarrow{p_3}\right) ,
\phi _2\left( \overrightarrow{p_3}\right) \right) _3+\left(
\varphi _1\left( \overrightarrow{p_4}\right) ,\varphi _2
\left( \overrightarrow{p_4}\right) \right) _4
\end{equation}
The set of kinetic equations describing a non-equilibrium state of the  two -component system of impuritons and phonons can be   represented in the operator form:

\begin{equation}\label{kin}I_{33}\left| g\right\rangle +I_{44}\left| g\right\rangle +I_{34}\left| g\right\rangle +L\left| g\right\rangle =\left| V\right\rangle
\end{equation}
Here 
{$\left| g\right\rangle =
\left| g_3; g_4\right\rangle ^T $,
$\left| V\right\rangle =\left| 
\overrightarrow{v_3}\overrightarrow{\nabla_3}; \overrightarrow{v_4}\overrightarrow{\nabla_4}
\right\rangle ^T$},
$\overrightarrow{\nabla_3}=\frac{\partial\left(\frac{H_3-\mu_3}T\right)}{%
\partial \overrightarrow{r}}$,
$\overrightarrow{\nabla_4}=\frac{\partial
\left(\frac{H_4}T\right) }{\partial \overrightarrow{r}}$,
$\overrightarrow{v_i}=
\frac{\partial \varepsilon _i}{\partial \overrightarrow{p_i}}$ are  the velocities of  quasiparticles   of species $i$  in the s-system, $f_i$  are the quasiparticle distribution functions, $H_3=\varepsilon _0+\varepsilon _3+\overrightarrow{p_3}\overrightarrow{v_s}$,
$H_4=\varepsilon_4+\overrightarrow{p_4}\overrightarrow{v_s}$  are the Hamiltonians  of the impuritons and phonons respectively, $\overrightarrow{v_s}$  is the  velocity of the  superfluid component, $g_i=\left( f_0^{\left( i\right) ^{\prime }}\left( \frac{H_i}T\right)\right)^{-1}\left( f_i-f_0^{\left( i\right)}\right)$  are the  small  corrections to  the equilibrium distribution functions $f_0^{\left( i\right)}$. 
We defined the  collision operators in equation ~(\ref{kin})   as:

\mbox{\indent $
I_{33}=\left( 
\begin{array}{cc}
\widehat{I}_{33} & 0 \\ 
0 & 0 
\end{array}
\right) 
$;\indent
$
I_{44}=\left( 
\begin{array}{cc}
0 & 0 \\ 
0 & \widehat{I}_{44} 
\end{array}
\right) 
$;\indent
$I_{34}=\left( 
\begin{array}{cc}
\widehat{I}_{34}^{\left( 3\right) } & \widehat{I}_{34}^{\left( 4\right) } \\ 
\widehat{I}_{43}^{\left( 3\right) } & \widehat{I}_{43}^{\left( 4\right) } 
\end{array}
\right) $;
\indent
$L=\left( 
\begin{array}{cc}
\widehat{L}_3 & 0 \\ 
0 & \widehat{L}_4 
\end{array}
\right)$}

\begin{equation}
\label{3}
\begin{array}{c}	
\widehat{I}_{ii}g_i=\int W_{ii}\left( \overrightarrow{p_i},\overrightarrow{p_{i_1}}\right| \left. \overrightarrow{p_{i_2}},\overrightarrow{p_{i_3}}\right) \left( 1\pm f_0^{\left( i\right) }\left( p_i\right) \right) ^{-1}\left( 1\pm f_0^{\left( i\right) }\left( p_{i_2}\right) \right) f_0^{\left( i\right) }\left( p_{i_1}\right) \left( 1\pm f_0^{\left( i\right) }\left( p_{i_3}\right) \right)\\ 
\left[ g_i\left( \overrightarrow{p_{i_2}}\right) +g_i\left( \overrightarrow{p_{i_3}}\right) -\right. \left. g_i\left( \overrightarrow{p_i}\right) -g_i\left( \overrightarrow{p_{i_1}}\right) \right] d\Gamma _{i_1}d\Gamma _{i_2}d\Gamma _{i_3}
\end{array}
\end{equation}

\begin{equation}
\label{5}
\begin{array}{c}
\widehat{I}_{ij}^{\left( i\right) } g_i=\int W_{34}\left( \overrightarrow{p_i},\overrightarrow{p_j}\right| \left. \overrightarrow{p_{i_1}},\overrightarrow{p_{j_1}}\right) \left( 1\pm f_0^{\left( i\right) }\left( p_i\right) \right) ^{-1}\left( 1\pm f_0^{\left( i\right) }\left( p_{i_1}\right) \right) f_0^{\left( j\right) }\left( p_j\right) \left( 1\pm f_0^{\left( j\right)} 
\left({p_{j_1}} \right) \right) \\
\left[ g_i\left( \overrightarrow{p_{i_1}}\right) -g_i\left( \overrightarrow{p_i}\right) \right] d\Gamma _{i_1}d\Gamma _jd\Gamma _{j_1}
\end{array}
\end{equation}

\begin{equation}
\label{1}	
\begin{array}{c}
\widehat{I}_{ij}^{\left( j\right) } g_j=\int W_{34}\left( \overrightarrow{p_i},\overrightarrow{p_j}\right| \left. \overrightarrow{p_{i_1}},\overrightarrow{p_{j_1}}\right) \left( 1\pm f_0^{\left( i\right) }\left( p_i\right) \right) ^{-1}\left( 1\pm f_0^{\left( i\right) }\left( p_{i_1}\right) \right) f_0^{\left( j\right) }\left( p_j\right) \left( 1\pm f_0^{\left( j\right)} 
\left({p_{j_1}}\right) \right) \\
\left[ g_j\left( \overrightarrow{p_{j_1}}\right) -g_j\left( \overrightarrow{p_j}\right) \right] d\Gamma _{i_1}d\Gamma _jd\Gamma _{j_1}
\end{array}
\end{equation}
Here $\widehat{I}_{ii}$ are the linearized impuriton-impuriton ($i=3$) and phonon-phonon ($i=4$) collision operators ~[\ref{11},\ref{Adam}], $\widehat{I}_{ij}^{\left( i\right) }$ and $\widehat{I}_{ij}^{\left( j\right) }$  are the components of the  linearized operators of cross impuriton-phonon  collisions which act in the spaces $\Re _i$, $\Re _j$ respectively, $L$ is the operator describing   the  quasiparticle-macron collisions, sign "+" corresponds to phonons,  "-" to  impuritons.
	
	The procedure of solving of kinetic equation ~(\ref{kin}) is simplified considerably if one uses the Lorentz approximation [~\ref{2}] for the operator $L$   describing the scattering of the "light" quasiparticles by the "heavy" macrons. The components of  this  operator describing the  impuriton-macron ($i=3$)  and phonon-macron ($i=4$) collisions   can be presented in the form:

\begin{equation}
\label{Ldef}
\widehat{L}_i g_i=N \int{ W\left( \overrightarrow{p_i}\right| \left. \overrightarrow{
p_{i_1}}\right) \left( g\left( \overrightarrow{p_{i_1}}\right)
-g\left( \overrightarrow{p_i}\right) \right) \delta \left( \varepsilon _i
\left( p_i\right) -\varepsilon _i\left( p_{i_1}\right) +
\overrightarrow{v}_s\left( \overrightarrow{p_i}-\overrightarrow{p_{i_1}}\right) \right) d\Gamma_{i_1}}
\end{equation}
where  $N$  is the macron  number density. 

The   "light" particles  diffusing in a "heavy" gas change only the direction of their motion due to the  collisions so that their energy remains unchanged in the system where the scatterers are at rest  ~[\ref{2}]. It is similar to a  gas of long-wavelength phonons which are scattered by macrons
 ~[\ref{2},\ref{9}]. Thus, every  scattering of a quasiparticle by the macron leads to a change in the  quasiparticle energy  which is equal to
$\overrightarrow{v}_s\left( \overrightarrow{p_i}-\overrightarrow{p_{i_1}}\right) $
in the s-system (see ~(\ref{Ldef})). It should be noted that the corresponding change  of  energy in the  process of the quasiparticles collisions with one another goes to  zero in the s-system due to momentum conservation law. The energy conservation  law in the  collisions of quasiparticles with macrons is taken into account  by the $\delta $ -functional factor  in collision integral (\ref{Ldef}). One can neglect   the term  proportional to $\overrightarrow{v_s}$  in the argument of  $\delta $ - function to the first order  in  the  thermodynamic gradients.  Thus,  the collision operators $\widehat{L}_i$  act only on the  polar and azimuthal  components of the  function of the  corresponding quasiparticle momentum. 

	In order to inverse the collision operator and solve operator equation ~(\ref{kin}) we  introduce the orthonormal basis  in the Hilbert space $\Re $, according to the definitions  :

{$
\left| \psi _{100}\right\rangle =\frac 1{\sqrt{4\pi }}\left|\begin{array}{c}
 1 \\ 1\end{array}\right\rangle
$;
$
\left| \psi _{200}\right\rangle =\frac 1{\sqrt{4\pi }}\left|\begin{array}{c}
\tau B_1 \\ -\frac{F_{-\frac 12}}2\end{array}\right\rangle
$;
$
\left| \psi _{300}\right\rangle =\frac 1{\sqrt{4\pi }}\left|\begin{array}{c}
p^2 \\ q\end{array}\right\rangle
$;
$
\left| \psi _{400}\right\rangle =\frac 1{\sqrt{4\pi }}\left|\begin{array}{c}
        4\tau \pi ^2B_2p^2 \\ -\frac 52F_{\frac 32}q\end{array}\right\rangle
$;} 
\mbox{$
\left| \psi _{110}\right\rangle =-i\sqrt{\frac{4\pi }3}\left|\begin{array}{c}
 Y_{10}\left( \theta \right) p \\ Y_{10}\left( \vartheta \right) q\end{array}
 \right\rangle
$;
$
\left| \psi _{210}\right\rangle =-i\sqrt{\frac{4\pi }3}\left|\begin{array}{c}
\frac 43\tau \pi ^2B_2Y_{10}\left( \theta \right) p \\ -\frac 12F_{\frac 12}Y_{10}
\left( \vartheta \right) q\end{array}\right\rangle
$;} \par
$$
\left| \psi _{11\pm 1}\right\rangle =\sqrt{\frac{2\pi }3}\left| 
\begin{array}{c}
\left( Y_{11}\left( \theta ,\varphi \right) \pm Y_{1-1}\left( \theta ,\varphi
\right) \right) p \\ \left( Y_{11}\left( \vartheta ,\phi \right) \pm Y_{1-1}
\left( \vartheta ,\phi \right) \right) q
\end{array} \right\rangle ;
$$
$$
\left| \psi _{21\pm 1}\right\rangle =i\sqrt{\frac{2\pi }3}\left| \begin{array}{c}
\frac 43\tau \pi ^2B_2\left( Y_{11}\left( \theta ,\varphi \right) \pm Y_{1-1}
\left( \theta ,\varphi \right) \right) p \\ -\frac 12F_{\frac 12}\left( Y_{11}
\left( \vartheta ,\phi \right) \pm Y_{1-1}\left( \vartheta ,\phi \right) \right) q
\end{array} \right\rangle 
$$
and

{$
\left| \psi _{2klm}\right\rangle =\left| \begin{array}{c} 0 \\ Y_{lm}
\left( \vartheta ,\phi \right) q^{k-1-\delta _l^1\delta _k^2}\end{array} \right\rangle
$;
$
\left| \psi _{2k-1lm}\right\rangle =\left| \begin{array}{c} Y_{lm}\left( \theta ,
\varphi \right) p^{k-1-\delta _l^0\delta _k^3-\delta _l^1\delta _k^2} \\ 0
\end{array} \right\rangle 
$\\
\mbox { where $k=3,4,5...$ for $l=0$,
 $k=2,3,4...$ for $l=1$,
 $k=1,2,3...$ for $l\geq 2$}.\par 
Here $Y_{lm}\left( \theta ,\varphi \right) $  are  the spherical functions~[\ref{10}], $\overrightarrow{p}=\frac{\overrightarrow{p_3}}{\sqrt{2m^{*}T}}$ and
$\overrightarrow{q}=\frac{c\overrightarrow{p_4}}T$  are the  dimensionless values of the  momentum of impuriton and phonon respectively, $\tau =
\left( \frac T{c\hbar }\right) ^3$, $B_n$  are  Bernoulli's   numbers, $F_\nu
=\left( \frac{\sqrt{2m^{*}T}}\hbar \right) ^3\frac 1{2\pi ^2}\int\limits_0^\infty \frac{x^{\nu}dx}{\exp(x-\frac{\mu_3}{T})+1}$ is the  Fermi-function, $\delta _m^l$ is the  Kronecker   symbol.
Let us note also  that the vectors introduced above satisfy  the conditions of orthonormality:

\begin{equation}
\label{ort}
\left\langle \psi _{2k-1lm}\right| \left. \psi _{2klm}\right\rangle=0
 \qquad  \mbox {k=1,2,3...;l=0,1,2...}
\end{equation}

Using the set of vectors, introduced above,  one can represent the conservation laws for  impuriton- phonon collisions  in the form:

\begin{equation}
\label{o34}		
I_{34}\left| \psi _{nml}\right\rangle =0
\end{equation}
where the set of indices $\left( n,m,l\right) $  corresponds to:

$\left( 1,0,0\right) $ -   number density conservation law;\par
$\left( 3,0,0\right) $ -  energy conservation law;\par
$\left( 1,1,0\right) $ -  z - component of momentum conservation law;\par
 $\left( 1,1,1\right) $ -  x- component of momentum conservation law;\par
$\left( 1,1,-1\right) $ -  y- component of momentum conservation law.\par

A  similar formula  describes the conservation laws in the collisions of quasiparticles of same species with one another:

\begin{equation}
\label{o33}
I\left| \psi _{nml}\right\rangle =0
\end{equation}
with the operator $I=I_{33}+I_{44}$ and  the  sets of indices to be  added to the above mentioned, as follows: $(2,0,0), (4,0,0), (2,1,0), (2,1,1), (2,1,-1)$. 

It should be noted that according to formulae ~(\ref{o34}), ~(\ref{o33})  the operators $I_{34}$ and  $I$  have  kernels which represent  subspaces of different dimensions.  It is concerned with the fact that the operator $I$   has ten independent collision invariants: five for impuritons collisions  with one another   and five for phonons ones. The  operator $I_{34}$ has only five invariants corresponding to number density, energy and momentum conservation laws  in  impuriton- phonon collisions. 

	The above  complete set of the  vectors can be orthonormalized by means of  the recurrent formulae : 

\begin{equation}
\label{fi1}
\left| \varphi _{0lm}\right\rangle =\frac{\left| \psi _{0lm}\right\rangle }{\sqrt{\left\langle \psi _{0lm}\right| \left. \psi _{0lm}\right\rangle }}
\end{equation}			
\begin{equation}
\label{fi2}
\left| \varphi _{n+1lm}\right\rangle =\frac{\left( E-P_0^{\left( n,l,m\right) }\left[ \left\{ \varphi _{klm}\right\} _{k=1}^n\right] \right) \left| \psi _{n+1lm}\right\rangle }{\sqrt{\left\langle \psi _{n+1lm}\right| \left. \psi _{n+1lm}\right\rangle -\left| \left| \psi _{n+1lm}\right| \right| _n^2}}
\end{equation}
where

\begin{equation}
\label{pr1}
P_0^{\left( n,l,m\right) }\left[ \left\{ \varphi _{klm}\right\} _{k=1}^n\right] =\sum\limits_{k=1}^n\left| \varphi _{klm}\right\rangle \left\langle \varphi _{klm}\right| 
\end{equation}

is  the set of projectors produced  by  the orthonormal basis $\left\{ \left| \varphi _{klm}\right\rangle \right\} $,
\begin{equation}
\label{nr}
\left| \left| \psi _{n+1lm}\right| \right| _n=\sqrt{\sum\limits_{k=1}^n\left\langle \varphi _{klm}\right| \left. \psi _{n+1lm}\right\rangle ^2}
\end{equation}
is  the norm of a projection of  the vector $\left| \psi _{n+1lm}\right\rangle$  to the subspace with the basis $\left\{ \varphi _{klm}\right\} _{k=1}^n$.

 The operators 
\begin{equation}
\label{pr34}
P_0^{\left[ 3,4\right] }=P_0^{\left( 2,0,0\right) }\left[ \left\{ \varphi _{2k-100}\right\} _{k=1}^2\right] +\sum\limits_{m=-1}^1P_0^{\left( 1,1,m\right) }\left[ \varphi _{11m}\right] 
\end{equation}

\begin{equation}
\label{pr0}
P_0=P_0^{\left( 4,0,0\right) }\left[ \left\{ \varphi _{k00}\right\} _{k=1}^4\right] +\sum\limits_{m=-1}^1P_0^{\left( 2,1,m\right) }\left[ \left\{ \varphi _{k1m}\right\} _{k=1}^2\right]
\end{equation}

represent  the projectors of Hilbert space $\Re $  to the kernels of operators   $I_{34}$  and $I$,  respectively. It is convenient to rewrite operator ~(\ref{pr0})  in the form of a  multiplication:

\begin{equation}
\label{pr2}
P_0=P_0^{\left[ 3,3\right] }P_0^{\left[ 4,4\right] }
\end{equation}

where

\begin{equation}
\label{pr33}
P_0^{\left[ 3,3\right] }=\sum\limits_{l=0}^\infty \sum\limits_{m=-l}^lP_0^{\left( \infty ,l,m\right) }\left[ \left\{ \varphi _{2k-100}\right\} _{k=1}^2,\left\{ \varphi _{11m}\right\} _{m=-1}^1,\left\{ \varphi _{2klm}\right\} _{k=1}^\infty\right]
\end{equation}

\begin{equation}
\label{pr44}
P_0^{\left[ 4,4\right] }=\sum\limits_{l=0}^\infty \sum\limits_{m=-l}^lP_0^{\left( \infty ,l,m\right) }\left[ \left\{ \varphi _{2k00}\right\} _{k=1}^2,\left\{ \varphi _{21m}\right\} _{m=-1}^1,\left\{ \varphi _{2k-1lm}\right\} _{k=1}^\infty\right]
\end{equation}
are  the projectors to the kernels of operators $I_{33}$ , $I_{44}$  describing the mutual collisions  of impuritons and phonons, respectively. Besides the operators ~(\ref{pr34}), ~(\ref{pr33}), ~(\ref{pr44}) we  introduce the projector $P_L$  to the kernel of operator $L$.  According to the definition of the operator $L$  and all the notes to formula ~(\ref{Ldef})  concerning it's kernel one can obtain:

\begin{equation}
\label{prL}
 P_L=P^{\left( \infty ,0,0\right) }
\end{equation}

The above introduced sets  of the  orthonormal vectors  and  the derived  projectors  make it possible to obtain the general solution of operator equation ~(\ref{kin}).  For this purpose let us project the latter to the ranges of values of operators $P_0^{\left[ 1\right] }=P_0^{\left[ 3,4\right] }P_L$, $P_0^{\left[ 2\right] }=\left( P_0-P_0^{\left[ 3,4\right] }\right) P_L$, $P_0^{\left[ 3\right] }=P_0^{\left[ 3,4\right] }-P_0^{\left[ 1\right] }$  and to their orthogonal complement. As a consequence one derives
 the set of equations  which can be written as follows: 

\begin{equation}
\label{sys1} 
\begin{array}{c}
\left(\left( E-P_0^{\left[ 3,3\right] }\right) I_{33}\left( E-P_0^{\left[ 3,3\right] }\right) \right. +\left( E-P_0^{\left[ 4,4\right] }\right) I_{44}\left( E-P_0^{\left[ 4,4\right] }\right) + \\ \left( E-P_0^{\left[ 3,4\right] }\right) I_{34}\left(
E-P_0^{\left[ 3,4\right] }\right) +\left. \left( E-P_L\right) L\left( E-P_L\right) \right) \left| g\right\rangle =\left( E-P_0^{\left[ 1\right] }\right) \left| V\right\rangle
\end{array}
\end{equation}
\begin{equation}
\label{sys2}
 P_0^{\left[ 3\right] }L\left( E-P_L\right) \left| g\right\rangle
 =P_0^{\left[ 3\right] }\left| V\right\rangle  
\end{equation}
\begin{equation}
\label{sys3}
 P_0^{\left[ 2\right] }I_{34}\left( E-P_0^{\left[ 3,4\right] }\right) \left| g\right\rangle =P_0^{\left[ 2\right] }\left| V\right\rangle
\end{equation}
\begin{equation}
\label{sys4}
 P_0^{\left[ 1\right]}\left| V\right\rangle =0
\end{equation}
Here $E$ is the identity operator.

Because $P_L \left| V \right\rangle =0$,  equality (\ref{sys4}) is fulfilled trivially  and can be ommited. Equality ~(\ref{sys3}) means that the vector $I_{34} \left| g\right\rangle$  has to belong to  kernel of the  projector $P_0^{\left[ 2\right] }$. 

It should be noted that the operators $\left( E-P_0^{\left[ i,j\right] }\right) I_{ij}\left( E-P_0^{\left[ i,j\right] }\right) $ ($i,j=3,4$) in equation ~(\ref{sys1}) represent the restrictions of the  collision operators $I_{ij}$  with respect to the orthogonal complements of  their kernels. Because all the collision operators are non-positive ones  the sum in the left-hand side of equation ~(\ref{sys1}) can  be inverted one-to-one . The latter condition makes it possible to solve the set of operator equations  (\ref{sys1}) -~(\ref{sys4})   for the components of a non-equilibrium term $\left| g\right\rangle $  to be represented in the form:

\begin{equation}
\label{g}
\left| g\right\rangle =\left| g_0\right\rangle +\left| g^{^{\prime }}\right\rangle 
\end{equation}
where $\left| g_0\right\rangle $  is an arbitrary vector which belongs to the range  of values of operator $P_0^{\left[ 1\right] }$ (i.e. the solution of homogeneous operator equation   corresponding to  kinetic equation ~(\ref{kin})), and

\begin{equation}
\label{g'}
\begin{array}{c}
\left| g^{^{\prime }}\right\rangle =\left( \left( E-P_0^{\left[ 3,3\right] }\right) I_{33}\left( E-P_0^{\left[ 3,3\right]}\right) +\left( E-P_0^{\left[ 4,4\right]}\right) I_{44}\left( E-P_0^{\left[ 4,4\right]}\right) +\right. \\
\left( E-P_0^{\left[ 3,4\right] }\right) I_{34}\left( E-P_0^{\left[ 3,4\right] }\right) +\left. \left( E-P_L\right) L\left( E-P_L\right) \right) ^{-1}\left( E-P_0^{\left[ 1\right] }\right) \left| V\right\rangle 
\end{array}
\end{equation}

The obtained solution of kinetic problem  ~(\ref{g})-(\ref{g'})  is exact and can be used for the  investigation of a steady non-equilibrium state of the two- component impuriton-phonon system  in the  presence of macrons. It takes into account  that the  collision operator represents  the sum of the operators with  kernels of different dimensions. It can  be elucidated  in the limiting cases  when the contribution of a certain type of collisions can be neglected.  In these cases solution  ~(\ref{g'}) leads to the correct limiting expressions under the condition that one omits the  corresponding collisions operator in it.  If  one can neglect the contribution of phonons formula ~(\ref{g'})    gives the result presented in ~[\ref{17}]. 

	If the contribution of mutual  collisions of both species  of quasiparticles should be taken into account  solution ~(\ref{g'}) can be represented    in the form:

\begin{equation}
\label{g'1}	
\begin{array}{c}
\left| g^{^{\prime }}\right\rangle =\left( \left( E-P_0\right) I\left( E-P_0\right) \right. +\left( E-P_0^{\left[ 3,4\right] }\right) I_{34}\left( E-P_0^{\left[ 3,4\right] }\right) +  \\
\left. \left( E-P_L\right) L\left( E-P_L\right) \right) ^{-1}\left( E-P_0^{\left[ 1\right] }\right) \left| V\right\rangle  
\end{array}
\end{equation}

It should be noted that  solution of the  kinetic equation ~(\ref{g'})  has to fulfill  conditions ~(\ref{sys2}) and  ~(\ref{sys3}). To illustrate it let us consider the hydrodynamic limit when the contribution of quasiparticle - macron collisions can be neglected. To obtain the correct limiting transition  in the  set of equations  ~(\ref{sys1}) -  ~(\ref{sys4}) one should   use the equality $P_L=E$. As a result  equation ~(\ref{sys2}) reduces to the form:

\begin{equation}
\label{eq}
P_0^{\left[ 3,4\right]}\left| V\right\rangle =0
\end{equation}

Equality ~(\ref{eq})  yields the condition of solvability   for equation ~(\ref{kin}) describing a  steady non-equilibrium state of the two-component phonon-impuriton system. Using the explicit expressions  for projector $P_0^{\left[ 3,4\right]}$  ~(\ref{o34}) and vector $\left| V\right\rangle $
 one can rewrite the condition ~(\ref{eq}) as follows:

\begin{equation}
\label{eq1}
\frac{\partial P_3}{\partial \overrightarrow{r}}+S_{ph}\frac{\partial T}{\partial \overrightarrow{r}}=0
\end{equation}
where $P_3=\frac 23TF_{\frac 32}$  is  the osmotic pressure of the gas of impuritons; $S_{ph}\left(T\right)$  is the entropy of phonons per unit volume and  $S_{ph}\frac{\partial T}{\partial \overrightarrow{r}}$ is the gradient of fountain pressure caused by phonons ~[\ref{177},\ref{4}]. Relation ~(\ref{eq1}) represents a conventional condition of a  mechanical equilibrium in a  closed system: the total pressure must be constant in entire volume of the mixture 
~[\ref{177}]. The remarkable fact is  that one can even  avoid  to use  solution ~(\ref{g})  of the kinetic equation to  obtain the condition describing a steady non-equilibrium state of the mixture in the hydrodynamic limit. For this purpose it is enough to analyze the condition of it's solvability ~(\ref{sys2}) which reflects  the momentum conservation law in impuriton - phonon collisions. 

	In the  presence of the  macrons the phonon-impuriton system is unclosed.   The operator $L$  is not equal to zero and  relation  ~(\ref{eq1}) is not fulfilled. So,  in order  to investigate the non-equilibrium state of the mixture one has to analyze solution ~(\ref{g}) of the kinetic equation.

\section{Futher analysis of the  solution of a kinetic equation by using the model representation of a collision operator}

The exact solution of the kinetic equation ~(\ref{g'1}) allows us to construct the approximations which   take  certain  properties of true collision operator into account. Here we restrict  ourselves to  the model for a two-component system collision operator which should be considered as a generalization of the approach developed in ~[\ref{Cherch},\ref{Mors}]  for the case of the one-component gas. 

The essential features  of the  true collision operators which is  reflected in the proposed  model  are  as follows: 

1) the model collision  operators  have to  satisfy  conditions ~(\ref{o34}), ~(\ref{o33})  expressing the conservation laws in  the  collisions;\par

2) linearized collision operators must be self-adjoint ;\par

3) the H- theorem ~[\ref{2}] must be satisfied.

According to the  first two conditions the collision operators must satisfy the relations:

\begin{equation}
\label{l33n}
I\left| g\right\rangle =\left( E-P_0\right) I\left( E-P_0\right)\left| g\right\rangle 
\end{equation}
\begin{equation}
\label{l34n}
I_{34}\left| g\right\rangle =\left( E-P_0^{\left[ 3,4\right] }\right) I_{34} \left( E-P_0^{\left[ 3,4\right] }\right) \left| g\right\rangle 
 \end{equation} 

	Developing the idea of the Bhatnagar-Gross-Krook approximation ~[\ref{Cherch}]  for  the  one-component gas it is natural to approximate the true collision operators $I$ and $I_{34}$ by the multiplication operators. Such an approach results in the model:

\begin{equation}
\label{mod}
I+I_{34}=\left( E-P_0\right) \widehat{\lambda }\left( E-P_0\right) +\left( E-P_0^{\left[ 3,4\right] }\right) \widehat{\lambda }_{34}\left( E-P_0^{\left[ 3,4\right] }\right) 
\end{equation}
where $\widehat{\lambda }=diag\left\{ -\nu _{33},-\nu _{44}\right\}$,
$ \widehat{\lambda }_{34}=diag\left\{ -\nu _{34},-\nu _{43}\right\} $
are the diagonal matrices with  real elements.
	It should be noted that the parameters $\nu _{ij}$$\left( i,j=3,4\right) $ must be positive to satisfy the conditions of H-theorem. To prove it let us consider the entropy of the  non-equilibrium phonon-impuriton system:

\begin{equation}
\label{S}
S=-\int \left( f_3\ln f_3+\left( 1-f_3\right) \ln \left( 1-f_3\right) \right) d\Gamma _3-\int \left( f_4\ln f_4-\left( 1+f_4\right) \ln \left( 1+f_4\right) \right) d\Gamma _4
\end{equation}

	Taking the time derivative of $S$  and linearizing the result with respect to the non-equilibrium term $\left| g\right\rangle $ we represent the condition of H-theorem in the form:
\begin{equation}
\label{H}
\left\langle g\right| I+I_{34}\left| g\right\rangle \leq 0
\end{equation}
Here, equality holds if (and only if) the vector $\left| g\right\rangle $ belongs to the kernel of operator $I+I_{34}$.
According to condition ~(\ref{H}) the model operators must be negative and, consequently, the above mentioned  parameters must satisfy  $\nu_{ij}>0$. 

The idea behind the representation of the model collision operator in the  form ~(\ref{mod}) is  that the average effect of the  collisions is to change the  distribution functions by  amounts proportional to the $\left| g_i\right\rangle$, i.e. by  the deviations of $f_i$ from their equilibrium values. The parameters $\nu_ {ij}$ play a role of coefficients at the non-equilibrium terms $\left| g_i\right\rangle$ which  should be associated with the collision frequencies of an according type. 

	Model ~(\ref{mod}) satisfies  the conditions 1)-3) for arbitrary positive values $\nu_{ij}$. But the above explained physical meaning  implies that the frequencies of impuriton-phonon ($\nu_{34}$)  and phonon-impuriton ($\nu_{43}$) collisions are connected with one another. In order to find this relation between the frequencies let us consider the corresponding true collision operators in details. The collision term can be   represented  in the form:

\begin{equation}
\label{W}
\widehat{I}_{ii}g_i+\widehat{I}_{ij}^{\left( i\right) }g_i+\widehat{I}_{ij}^{\left( j\right) }g_j=\sum_{j=3}^{4} \overline{w}_{ij}\left( \overrightarrow{p_i}\right) \left( M_{ij}-g_i\right)
\hspace{30 pt}
\mbox{$\left(i=3,4\right)$}
\end{equation}
where 

\begin{equation}
\label{Wij}
\begin{array}{c}
\overline{w}_{ij}\left( \overrightarrow{p_i}\right) =
\int W_{ij}\left( \overrightarrow{p_{i}},\overrightarrow{p_{j_1}}\right| \left. \overrightarrow{p_{i_2}},\overrightarrow{p_{j_3}}\right) \left( 1\pm f_0^{\left( i\right) }\left( p_{i}\right) \right) ^{-1}\left( 1\pm f_0^{\left( i\right) }\left( {p_{i_2}}\right) \right) f_0^{\left( j\right) }\left( p_{j_1}\right) \\
 \left( 1\pm f_0^{\left( j\right) }\left( p_{j_3}\right) \right) 
d\Gamma _{i_2}d\Gamma _{j_1}d\Gamma _{j_3}
\end{array}
\end{equation}

\begin{equation}
\label{M}
\begin{array}{c}
M_{ij}\left( \overrightarrow{p_i}\right) =
\int W_{ij}\left( \overrightarrow{p_i},\overrightarrow{p_{j_1}}\right| \left. \overrightarrow{p_{i_2}},\overrightarrow{p_{j_3}}\right) \left( 1\pm f_0^{\left( i\right) }\left( p_i\right) \right) ^{-1}\left( 1\pm f_0^{\left( i\right) }\left( p_{i_2}\right) \right) f_0^{\left( j\right) }\left( p_{j_1}\right) \left( 1\pm f_0^{\left( j\right) }\left( p_{j_3}\right) \right) \\
\left[ g_j\left( \overrightarrow{p_{j_3}}\right) -g_j\left( \overrightarrow{p_{j_1}}\right) +g_i\left( \overrightarrow{p_{i_2}}\right) \right]
 d\Gamma _{i_1}d\Gamma _jd\Gamma _{j_3}
\end{array}
\end{equation}
Here, sign "+" corresponds to phonons, "-"  to  impuritons.

Hence, according to definition  ~(\ref{Wij}) the frequencies $\overline{w}_{34}$,  $\overline{w}_{43}$ are connected with one another by the relation:

\begin{equation}
\label{rel}
\int \overline{w}_{34}f_0^{\left( 3\right) ^{\prime }}d\Gamma _3=\int \overline{w}_{43}f_0^{\left( 4\right) ^{\prime }}d\Gamma _4
\end{equation}

	Expression ~(\ref{W}) has the same structure as  model collision term ~(\ref{mod}). Indeed, the latter can be represented in the form:

\begin{equation}
\label{St}
\sum\limits_{j=3}^4\nu _{ij}\left( f_0^{\left( i\right) }\left( \frac{\varepsilon _i-p_{i_z}{V}_{ij}}T\right) -f_i\right)
\hspace{30 pt}
\mbox{$\left(i=3,4\right)$}
\end{equation}
where 
\begin{equation}
\label{V3}
{V}_{ii}=-T\frac{\left(p_{i_z},g_i\right) _i}{\left( p_{i_z},p_{i_z}\right) _i}
\hspace{30 pt}
\mbox{$\left(i=3,4\right)$}
\end{equation}
are the moduli of the mass velocities of the quasiparticles of species  $i$.
The index  $z$ denotes the projection of corresponding vector on the direction of gradients. The values ${V}_{34}$,  ${V}_{43}$ are determined one-to-one by a comparison of  formulae ~(\ref{St}) and  ~(\ref{mod}). Thus, the complicated emission terms in the  true collision integrals are replaced by the multiplication of a  Maxwellian by the corresponding collision frequency in model presentation  ~(\ref{mod}).   It should be noted that such an approximation looks appropriate for the  conditions where the individual distribution functions are not far away from their equilibrium values. If the dependencies of  $\overline{w}_{ij}$ with respect to the  momentum are smooth enough it can be approximated by their mean values, i.e.  the effective collision frequencies $\nu _{ij}$ in model ~(\ref{mod}).  These frequencies may be conveniently introduced as follows. Let us apply the second mean value theorem to the integrals in equality ~(\ref{rel}). It gives:

\begin{equation}
\label{rel1}
\nu _{34}\int f_0^{\left( 3\right) ^{\prime }}d\Gamma _3=\nu _{43}\int f_0^{\left( 4\right) ^{\prime }}d\Gamma _4
\end{equation}
where the mean values are defined by:
\begin{equation}
\label{av}
\nu _{ij}=\frac{\left( 1,\overline{w}_{ij}\right) _i}{\left( 1,1\right) _i}
\end{equation}

Equality ~(\ref{rel1}) reduces to well-known relation between the cross collision frequencies and the number densities of gases in their two-component mixture ~[\ref{Mors}] by substituting the classical Maxwell distribution functions into it. Substituting the classical distribution function of impuritons and the Bose-distribution function of phonons into  equality ~(\ref{rel1}) we obtain:
\begin{equation}
\label{relt}
n_3 \nu _{34}=1.37 n_{ph} \nu _{43}
\end{equation}
where $ n_{ph}$ is the number density of phonons and $ n_{3}$ is the number density of impuritons.

We  investigate now the operator $L$ describing the quasiparticle-macron collisions. Let us consider the  non-equilibrium state of a mixture which  is  caused by a non-uniformity of the thermodynamic values along  the  z-axis. Thus, the non-equilibrium term $\left| g^{^{\prime }}\right\rangle $  cannot be a function of the polar angles in the  momentum spaces of both impuritons and phonons. The result of the  action of the collision operator on the non-equilibrium term $\left| g\right\rangle$ can be written in the form:

\begin{equation}
\label{Lm}
L\left| g\right\rangle =-\sum\limits_{n,l}\widehat{\nu }_{_{Kn}}^l\left| \varphi _{nl0}\right\rangle \left\langle \varphi _{nl0}\right. \left| g\right\rangle 
\end{equation}
where

\begin{equation}
\label{Kn}
\widehat{\nu }_{_{Kn}}^l=diag\left\{ \nu _{_{3Kn}}^l,\nu _{4Kn}^l\right\} 
\end{equation}
\begin{equation}
\label{Kn1}
\nu _{_{iKn}}^l=2\pi Nv_T^{\left( i\right) }\int \left( 1-P_l\left( \cos \alpha \right) \right) \sigma _i\left( p_i,\alpha \right) \sin \alpha d\alpha 
\end{equation}
are the partial frequencies of the  collisions of quasiparticles of species  $i$ with the macrons,  $v_T^{\left( 3\right) }=\left( \frac{2T}{m^{*}}\right) ^{\frac 12}p$, $v_T^{\left( 4\right) }=c$, $\sigma _i\left( p_i,\alpha \right) \sin \alpha d\alpha$ are the differential scattering cross-section for the quasiparticles of species $i$;  $P_l\left( \cos \alpha \right) $ is  the Legendre polynomial of the $l$th order.

	When substituting expressions ~(\ref{mod}) and  ~(\ref{Lm})  into  the kinetic equation ~(\ref{kin}) one has to take into account that the model operators $I$ and $I_{34}$ present the operators of multiplication.  It means that the sum of operators $I+I_{34}+L$ maps into itself the  subspace of vectors ~(\ref{fi2}) corresponding to the $(l,0)$th harmonics. The latter  makes it possible to look for a solution of equation ~(\ref{kin}) in the subspace corresponding to the same harmonics as the vector in the left-hand side of kinetic equation ~(\ref{kin}) belongs to.
	
Inserting  ~(\ref{mod}) and  ~(\ref{Lm}) into  ~(\ref{kin}) we get:

\begin{equation}
\label{kin1}
\left( \widehat{\lambda }P_n+\left( P_n+P_c\right) \widehat{\lambda }_{34}\left( P_n+P_c\right) +\widehat{\nu }_{_{Kn}}^1\right) \left| g\right\rangle =\left| V\right\rangle
\end{equation}

with:

\begin{equation}
\label{prc}
P_c=\left| \varphi _{210}\right\rangle \left\langle \varphi _{210}\right| 
\end{equation}

\begin{equation}
\label{prn}
P_n=E-P_c-\left| \varphi _{110}\right\rangle \left\langle \varphi _{110}\right|
 \end{equation}

	Analyzing the  solution of the kinetic equation ~(\ref{kin1}) we restrict ourselves to  the case when the macrons can be regarded as rigid spheres of radius $a$. Thus the phonon-macron collision frequency $\nu _{_{4Kn}}^l$  which describes the Rayleigh scattering of the long-wavelength  phonons by rigid spherical macron in equation ~(\ref{kin1}), is represented in the form ~[\ref{G}]:
\begin{equation}
\label{4kn}
\nu _{_{4Kn}}^1=cN\left( \frac{4\pi }{3\hbar ^2}a^3\left( 1-\frac{\rho _k}\rho \right) \right) ^2q^4
\end{equation}
where $\rho $ is the density of a mixture, $\rho _k$ is the average  density of macron. The frequency of collisions  of "light" impuriton with the "heavy" macron can be  written as:

\begin{equation}
\label{3kn}
\nu _{_{3Kn}}^1=p\nu _{_{3Kn}}
\end{equation}
where $\nu _{_{3Kn}}=\pi Na^2\left( \frac{2T}{m^{*}}\right) ^{\frac 12}$.

In the case when frequency of phonon-macron collisions is much less than the sum of frequencies of phonon-impuriton and phonon-phonon ones  the contribution of thermal excitation to a size-effect in the gas of impuritons should be expected to be  maximal. Such a limit can be realized practically by choosing the appropriate parameters  in formula ~(\ref{4kn}). 	
In this limiting case  the procedure of inversion of the  collision operator and solution of kinetic equation~(\ref{kin1})  is  simplified significantly. In  zeroth order with respect to the  parameter $\frac{\nu _{_{4Kn}}^1}{\nu _{44}+\nu _{43}}$ the expression for the non-equilibrium terms $g_i$ can be written as follows:

\begin{equation}
\label{g3}
g_3=-\frac{ v_{3z}}{ \nu _3+\nu _{_{3Kn}}^1} \left( \nabla _3 -\frac{S_{ph}}{n_3}
\frac{\nabla T}{T} \right)- \frac{\gamma}{p+ \gamma} \frac {p_{3_z} V_{33}}{T}
\end{equation}

\begin{equation}
\label{g4}
g_4=- \frac{1}{\nu_4} \left(v_{4z} \nabla _4 + \frac{p_{4_z} c^2}{T}
\frac{\nabla T}{T} \right)-  \frac {p_{4_z} V_{44}}{T}
\end{equation}

where
$\gamma\left( n_3, T \right) =\frac{\nu _3}{\nu _{_{3Kn}}}$, 	
$\nu _3=\nu _{33}+\nu _{34}$, 
$\nu _4=\nu _{44}+\nu _{43}$.

	Using definitions ~(\ref{V3}) one can obtain the explicit expressions for the mass velocities of impuritons and phonons

\begin{equation}
\label{V33}
\overrightarrow{V}_{33}=-{\frac{1}{{n_3}{ m^{*}}{ \nu_{3Kn}}} \frac{G_0 \left( \gamma \right)}{G_1 \left(\gamma \right)} \left( T \nabla n_3 + n_3 \eta \left( \gamma \right) \nabla{T} + S_{ph} \nabla{T} \right)}
\end{equation}

\begin{equation}
\label{V44}
\overrightarrow{V}_{44}=-{\frac{1}{{n_3}{ m^{*}}{ \nu_{3Kn}}} \frac{G_0 \left( \gamma \right)}{G_1 \left(\gamma \right)} \left( T \nabla n_3 + n_3 \eta \left( \gamma \right) \nabla{T} \right)
- \left( \frac{ G_0 \left( \gamma \right)}{{\nu_{3Kn}} {G_1\left(\gamma \right)}} - 
\frac{\left(1+ \delta \right)^2}{\nu_{34}+\nu_{43}\delta}\right) \frac{S_{ph} \nabla T}{n_3 {m^*}}}
\end{equation}

respectively.

 Here,
$G_n\left( \gamma \right) =\frac{4}{3\sqrt{\pi}}  \int_0^{\infty}  {\frac{y^{\frac {n+3}{2}}}{\gamma + \sqrt{y}}dy}$,  $\delta= \frac{n_3 m^{*}}{\rho_{nph}}$,
  $\rho _{nph}$  is  the  normal density of phonons and
\begin{equation}
\label{eta}
\eta \left( \gamma \right) =\frac{G_2\left( \gamma \right) }{G_0\left( \gamma \right) }-\frac {3}{2}
\end{equation}

	The model proposed in this section makes it possible to obtain the explicit results without accounting a large amount of details of the inter-quasiparticles interaction, assuming that it is not likely to influence   the experimentally measured values significantly. The  results (\ref{g3}) - (\ref{V44}) which were obtained  from  model ~(\ref{mod}) are used  to analyze the steady non-equilibrium state of a superfluid mixture in confined geometry.

\section{Steady non-equilibrium state of the superfluid mixture of helium isotopes.}

There are two different stages in the process of the  formation  of the  steady non-equilibrium state of the  superfluid mixture in confined geometry.  During the first stage the superfluid component which does not encounter any resistance, overflows  rapidly to ensure the condition under which the chemical potential of $^{4}He$   is constant  all over the mixture ~[\ref{177}]. At this stage the pressure gradient in the mixture  is equal  to the sum of the  gradients of the  osmotic and the  fountain pressures. 

The second prolonged stage corresponds to the steady state formation  in the quasiparticles system of a superfluid mixture. During this stage the slow impuriton flow induced by the unbalanced gradients of temperature and concentration emerges in the system. 

The actual steady state sets in only as a result of a second stage, when the concentration gradient balances the temperature gradient  so that the mass velocities  of 
both the  impuriton gas  and  the mixture  are  equal to zero. Evidently, the duration of the second stage depends essentially on the amount  of the quasiparticle-macron collision contribution  to the process of a steady state formation.

To analyze the described processes quantitatively  let us write the above mentioned conditions of vanishing of impuritons and mixture mass flow in the form

\begin{equation}
\label{V333}
\overrightarrow{V}_{33}=0
\end{equation}

\begin{equation}
\label{ro}
\rho _s\overrightarrow{v}_s+\rho _{nph}\overrightarrow{V}_{44}+n_3m^{*}\overrightarrow{V}_{33}=0
\end{equation}
respectively. 
Here $\rho _s$ is the density of the superfluid component of the mixture.  

The explicit form of condition
 ~(\ref{V333}) represents the relation between the temperature gradient and the concentration gradient  ensuring the steady state of a mixture. This relation can be investigated with any degree of accuracy by using  definition
 ~(\ref{V3}) and the  general solution of the kinetic equation ~(\ref{g'}). Relation ~(\ref{ro}) defines the velocity of a superfluid motion which compensates the normal one to ensure  the steady state of a mixture. 

In the region of degeneracy the terms of the sum in expression ~(\ref{V33}) corresponding to the contribution of thermal excitations can be neglected. 
So,  in this case relation  ~(\ref{V333})  transforms into the result presented in ~[\ref{17}], describing the steady non-equilibrium state of a superfluid mixture in the ultra-low temperature region.

Thus, to study the contribution of phonons to the process of the  steady state formation in the superfluid mixture one can restrict  oneself to  the  case  of a classical mixture. Proceeding from the  explicit expression ~(\ref{V33}) for the mass velocity of the impuritons and the condition of stationarity ~(\ref{V333}) we get:

\begin{equation}
\label{stac}
n_3\eta \left( \gamma \right) \nabla T+T\nabla n_3+S_{ph}\nabla T=0
\end{equation}

For further analysis relation ~(\ref{stac}) can be conveniently expressed in  a coordinate independent-form  as a  function $\Psi $ of the  impuriton  number density and the  temperature:

\begin{equation}
\label{psi}
\nabla \Psi \left( n_3, T \right) =0
\end{equation}

Integrating  the  equality  ~(\ref{stac}) one can obtain the explicit expression for the function $\Psi $:

\begin{equation}
\label{psi1}
\Psi =n_3 {\theta}^{n\left(\theta, \gamma_{\lambda} \right)} -  \int_{\theta}^1 S_{ph}\left( t T_{\lambda} \right)t^{n\left( t, \gamma_{\lambda}  \right)}\frac{dt}t
\end{equation}
where $\theta=\frac{T}T_{\lambda}$  is the  dimensionless temperature of a  mixture, $ T_{\lambda}$ is the temperature of the  $\lambda$-transition,  $\gamma_{\lambda}=\gamma\left(n_3, T_{\lambda} \right)$ and
\begin{equation}
\label{n}
n\left( \theta, \gamma_{\lambda} \right)=-\frac{1}{6.5\ln{\theta}} \int_{\theta^{6.5}}^1 \eta \left( \gamma_{\lambda} t \right)\frac{dt}t
\end{equation}
It should  also be  noted that the function $\Psi$ is defined up to an arbitrary constant.

Relation ~(\ref{relt}), the actual dependence of the  pnonon-impuriton colission frequency on the temperature $\nu_{34} \sim {T^7}$   ~[\ref{Rud}] and  the fact that in classical temperature region the  frequency of  impuriton-impuriton collissions  is much less than the frequency of impuriton-phonon ones ~[\ref{Rud}] have been taken  into account when deriving formula ~(\ref{psi1}).

Results (\ref{stac}) and  (\ref{psi}) describe the steady, thermodynamically non-equilibrium state of the  phonon-imputiton system for arbitrary relations between the frequencies of impuriton-impuriton, impuriton-phonon and impuriton-macron collisions. They allow to investigate all  cases  which can be  realized in experiments. 
In the hydrodynamic limit ($\gamma \rightarrow \infty $), when the contribution of impuriton-macron collisions can be neglected in comparison with both impuriton-impuriton and impuriton-phonon collisions, result ~(\ref{stac}) transforms to  relation ~(\ref{eq1}). In the  opposite, Knudsen limiting case ($\gamma =0$) relation ~(\ref{psi}) gives:

\begin{equation}
\label{knf}
\nabla \left( n_3\sqrt{T}+\frac 27S_{ph}\sqrt{T}\right) =0
\end{equation}

Relation (\ref{knf}) describes the Knudsen effect in the gas of impuritons of a  superfluid mixture of helium isotopes by taking the contribution of phonons  into consideration. The existence of a second term in the sum  in the parentheses in formula (\ref{knf}) makes the difference with famous Knudsen relation ~[\ref{2}]. The relative contributions of the "impuriton" and "phonon" terms to the sum in (\ref{knf}) are determined by the relation between the normal densities of impuritons and phonons. Hence, it should be expected that relation ~(\ref{knf}) deviates significantly from the classical Knudsen one for dilute mixtures at classical temperatures.

Result  (\ref{knf}) can be checked experimentally  by  the conventional  setup which is  used for osmotic pressure measurements ~[\ref{Ebner}]. It represents  the system of two containers filled with the superfluid mixture and  connected through the supergap   containing a porous material. The containers are maintained at  constant but different temperatures so that the  unbalanced gradients of temperature and concentration lead to an overflow of the mixture through the supergap. The  experiments ~[\ref{Ebner}] focus on the  the difference in pressures of the  mixture in containers which is formed  as a result of the rapid overflow of the  superfluid component. An increase in the expectation time in the experiments  similar to those described in ~[\ref{Ebner}] must  result in a realization of the above-mentioned second prolonged stage of a  steady state formation.   An experiment like this would  make it possible to analyze the Knudsen effect in the superfluid mixture described by formula ~(\ref{knf}). 

To investigate the  condition of stationarity ~(\ref{V333}) in an intermediate region one has to use relation ~(\ref{psi}) with function $\Psi $ ~(\ref{psi1}). The parameter $n$($\frac 12\leq n\leq 1$) introduced in  ~(\ref{psi1}) should be considered as an index   characterizing  the regime which was formed in the impuritons system of a superfluid mixture. The value $n=\frac 12$ corresponds to Knudsen limit, $n=1$  to the  hydrodynamic one. The intermediate region with  arbitrary relations between inter-quasiparticles and impuriton-macron collisions frequencies was investigated numerically. The result for $T=0.7 K$ is shown on the Fig.~[1]. As it can be seen the hydrodynamic regime is established in the system exponentially fast  as parameter $\gamma_{\lambda}$ increases.

If one can neglect the temperature dependence of the parameter $n$  a  simple approximation for the  function  $\Psi $  can be derived from ~(\ref{psi1}) to rewrite the  condition of stationarity ~(\ref{psi}):

\begin{equation}
\label{psi2}
\nabla \left( n_3T^n+\frac{S_{ph}T^n}{n+3}\right)=0
\end{equation}

Formula ~(\ref{psi2}) describes explicitly the transition from the Knudsen regime $\left(n=\frac{1}{2}\right)$ to the hydrodynamic one  $\left(n=1 \right)$.

There is an explicit physical explanation behind the  deviation of conditions ~(\ref{stac}) and (\ref{knf}) describing the steady non-equilibrium state  of a superfluid mixture  from the classical Knudsen relation. The requirement of vanishing  flow of an impuriton gas in form ~(\ref{V333}) describes the situation when the diffusion processes caused by the temperature gradient  and the  concentration gradient are equilibrated. According to ~(\ref{V33})  the flow of phonons caused by the gradient of temperature contributes to the coefficient of thermal diffusion of impuritons due to phonon-impuriton collisions. Hence, the existence of the phonon gas flow must lead to an   effective increase of the concentration gradient to equilibrate the process of thermal diffusion of impuritons while satisfying condition ~(\ref{V333}). Thus, the gradient of impuriton concentration  must  be larger in the presence of phonons  than in the case of  the classical Knudsen effect at the same conditions.

\section{Conclusions}

The size effect in a superfluid mixture of helium isotopes was investigated by taking  the contribution of thermal excitations into consideration. The  exact solution  ~(\ref{g}),~(\ref{g'}) of set of the kinetic equations ~(\ref{kin}) describing the steady, thermodynamically non-equilibrium state of a phonon-impuriton system has been obtained. The problem of the  steady flow of the superfluid mixture caused by the thermodynamic gradients  through the volume filled with the macrons was solved. Conditions  ~(\ref{V333}) and  ~(\ref{ro}) ensuring the steady  state of a superfluid mixture  in the volume filled with macrons were obtained and analyzed  explicitly.
 
A  model  ~(\ref{mod}) reflecting the significant features of  the true  collision operator   was proposed. It  made  it possible to simplify considerably  the main results obtained within the  considered approximation. In particular, the explicit results for the mass velocities ~(\ref{V33}) and  ~(\ref{V44}) of quasiparticle gas flow were obtained and  used in conditions ~(\ref{V333}) and  ~(\ref{ro}) which  describe the steady non-equilibrium state of a superfluid mixture in the volume filled with macrons. As a result, the  conditions of stationarity ~(\ref{V333}) and  ~(\ref{ro}) were  presented in the form of relation between the thermodynamic gradients ensuring the steady state of a mixture. The relation obtained was investigated  in the region of classical temperatures where the contribution of   phonons  is maximal. All the results are available for arbitrary relations between the frequencies of collisions of quasiparticles with one  another and with  macrons. The  steady state condition ~(\ref{stac})  was studied for arbitrary  relations between the frequencies of collisions of different types. 

The  hydrodynamic ($\nu_3>>\nu_{3Kn}$) ~(\ref{eq1}) and Knudsen ($\nu_3<<\nu_{3Kn}$) ~(\ref{knf}) limits of  the conditions ensuring  a steady state of a superfluid mixture   were investigated in details by taking  the contribution of phonons into account.  In particular, the correction to  the classical Knudsen condition caused by the presence of phonons was obtained. A convenient interpolation formula  ~(\ref{psi2}) was derived from the exact result  to observe explicitly  the steady state condition  ~(\ref{psi}) in the intermediate region $\nu_3 \sim \nu_{3Kn}$.

 The results  were  applied to the  investigations of a steady non-equilibrium state of a superfluid mixture situated in the volume filled with the macroparticles. Two different stages in the process of a steady state formation  were found  and studied. The experiment for an  investigation of the Knudsen effect in a superfluid mixture was proposed.

 \section{Acknowledgements}
Many thanks to professor Igor N. Adamenko for the  fruitful idea which initiated the series of works prolonged by this paper.
I am grateful   to the G\"asteprogramm of Max-Planck-Institut f\"ur Physik komplexer Systeme and to the  organizers of the seminar '' Topological Defects in Non-Equilibrium Systems and Condensed Matter''  for  possibility to complete and present this  work.
I would like to thank to Dr. Dieter Joseph for useful advise helped   in  presentation of  the    material.

\begin{figure}
\label{pic}
\centerline{\psfig{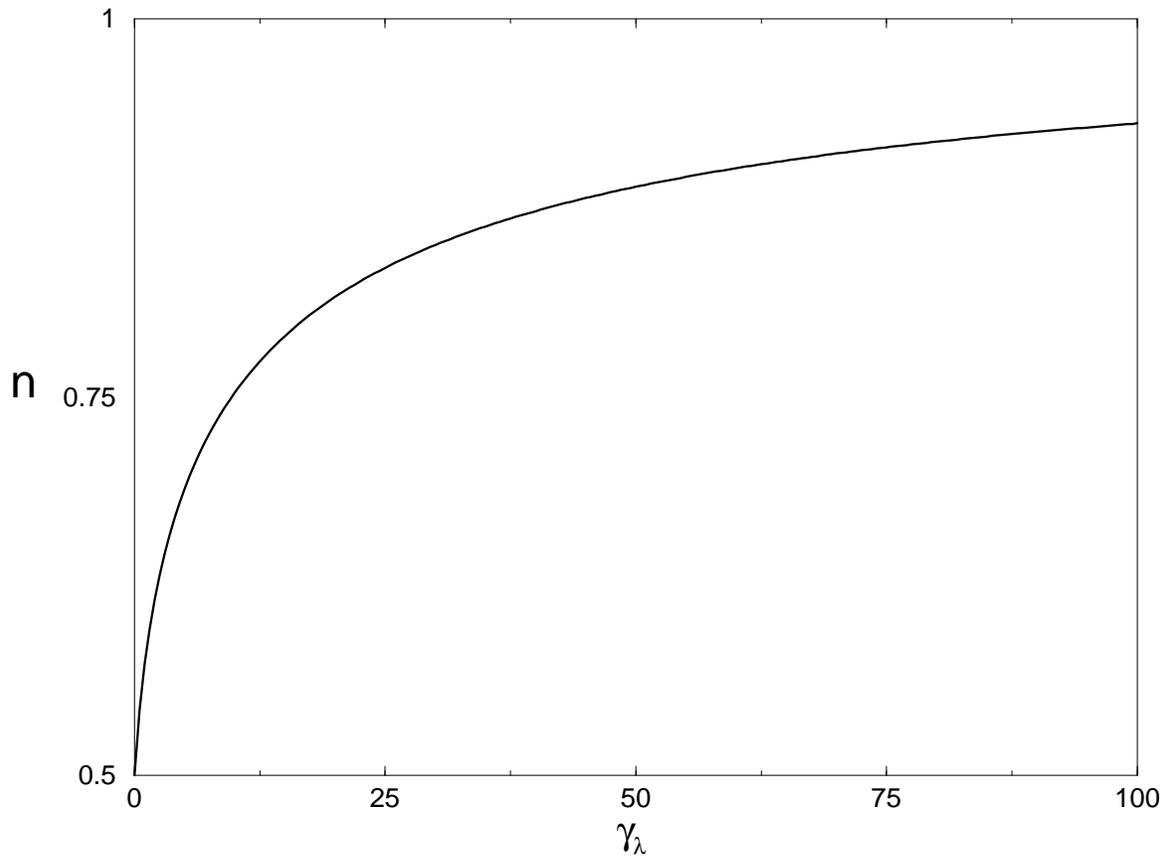}}
\vspace{0.5cm} 
\caption{The transition from the Knudsen regime to hydrodynamical one in relation ~(\ref{psi1}) describing the sufficient condition ensuring a steady state of the superfluid mixture $\left( T=0.7 K \right)$.} 
\end{figure}

\end{document}